\documentclass[fleqn,useAMS,usenatbib]{mnras}
\usepackage{mathptmx}
\usepackage{color}
\usepackage{graphics,graphicx,amssymb,amsmath,natbib}
\usepackage{float}
\usepackage{color}
\usepackage{lscape,graphicx}
\usepackage{amsmath}
\usepackage{amssymb}
\usepackage{multirow}
\usepackage{afterpage}
\usepackage{threeparttable}
\usepackage{subfigure}
\usepackage{rotating}
\usepackage{lscape}
\usepackage{caption}
\usepackage{verbatim}
\usepackage{morefloats}
\usepackage[T1]{fontenc}
\usepackage{ae,aecompl}
\usepackage{graphicx}
\usepackage{epstopdf}
\usepackage{amsmath}
\usepackage{amssymb}
\usepackage{enumerate}
\usepackage{hyperref}
\title[A Diffusion model for HESS~J1825-137]{On the Unusually Large Spatial Extent of the TeV nebula HESS~J1825-137: Implication from the Energy-Dependent Morphology}

\author[Liu and Yan]{Ruo-Yu Liu$^{1,2}$\thanks{E-mail: ryliu@nju.edu.cn}, Huirong Yan$^{2,3}$\\
$^1$School of Astronomy and Space Science, Nanjing University, Xian Lin Da Dao 163, 210023 Nanjing, China\\
$^2$Deutsches Elektronen Synchrotron (DESY), Platanenallee 6, D-15738 Zeuthen, Germany\\
$^3$Institut f\"ur Physik und Astronomie, Universit\"at Potsdam, D-14476 Potsdam, Germany}

\begin{document}

\maketitle

\begin{abstract}
Deep observation of the High Energy Stereoscopic System (HESS) on the most extended pulsar wind nebula HESS~J1825-137 reveals an enhanced energy-dependent morphology, providing useful information on the particle transport mechanism in the nebula. We find that the energy-dependent morphology is consistent with a diffusion-dominated transport of electrons/positrons. It provides an alternative possible interpretation for the unusually large spatial extent (i.e., $\gtrsim 100\,$pc) of the nebula, which could then be attributed to the diffusion of escaping electrons/positrons from a compact plerion. The influence of various model parameters on the energy-dependent extent of the nebula is studied in the diffusion-dominated scenario. {We also show that the energy-dependent morphology of the nebula may also be used to study the spin-down history of the pulsar.}
\end{abstract}


\section{Introduction}
Pulsar wind nebulae (PWNe) constitute one of the largest source population at very high energies (VHE, $>0.1$\,TeV). A PWN is powered by the associated pulsar through converting its spin-down energy to the nonthermal energy of electron/positron pairs \citep[e.g.][]{GS06}, {somewhere around the strong termination shock \citep[e.g.][]{KC84, Lyubarsky03, Amato06, Kirk09, Sironi11}} formed by the interaction between the ultrarelativistic pulsar wind and the surrounding medium. The VHE emission is believed to arise from TeV electrons (for simplicity, we do not distinguish positrons from electrons hereafter) by inverse Compton (IC) scattering on background photon field such as cosmic microwave background (CMB) and interstellar radiation field \citep[e.g.][]{Slane17}.
Due to the extended nature and their proximity to Earth, many PWNe are spatially resolved and hence serve as natural laboratories for studies of some fundamental processes such as acceleration and transport of ultrarelativistic particles, and eventually may provide a clue to the origin of high-energy cosmic rays.  

Among numerous observed PWNe, HESS~J1825-137 is one of the most luminous and the most extended PWN. It is associated with an energetic pulsar PSR~J1826-1334 (also known as PSR~B1823-13) which approximately locates at 4\,kpc away from Earth according to the dispersion measure of the pulsar \citep{Taylor93, Manchester05}. Given the rotation period $P=101.3\,$ms and the period derivative $\dot{P}=7.5\times 10^{-14}$ \citep{Clifton92}, the spin-down luminosity of PSR~J1826-1334 at the present time is $L_s=2.83\times 10^{36}\,$erg/s with a characteristic age of $\tau_c\equiv P/2\dot{P}=21.4\,$kyr. The High Energy Stereoscopic System (HESS) has revealed the diffuse VHE emission $>100\,$GeV extending from the pulsar \citep{HESS06_PWN} out to $\sim 1.5^\circ$ towards the south of PSR~J1826-1334 corresponding to a projected size of 100\,pc, while the emission steeply decreases towards the north of the pulsar. Such a north-south asymmetry was also observed in the X-ray band \citep{Pavlov08, Uchiyama09} and was attributed to the crushing effect of an asymmetric reverse shock arising from the supernova shell on the northern side \citep{HESS06_PWN}. Such an interpretation is supported by the presence of a dense molecular cloud north of the nebula \citep{Castelletti12, Voisin16}.

The formation of the unusually large spatial extension of the southern side of the TeV nebula has been studied by different authors, by considering the hydrodynamic evolution of the PWN inside a supernova remnant. \citet{dejager09} suggested that a low density of of $10^{-3}\rm cm^{-3}$ for the interstellar medium (ISM) around the supernova remnant is needed to reproduce the large size. \citet{Khangulyan18} pointed out that such a requirement may be inconsistent with the presence of the dense molecular cloud, and they proposed an alternative scenario considering a huge amount injection of the kinetic energy from the pulsar wind with a typical ISM density (e.g., $1\rm cm^{-3}$) for the ambient medium. This scenario could be achieved if the pulsar was born with a very short rotation period, i.e., $1\,$ms with a small braking index $n\leq 2$. In these studies, the extended TeV nebula HESS~J1825-137 is considered to be a plerion, namely, the TeV-emitting electrons are assumed to be well confined in the post-shock pulsar wind and are advected out to $100\,$pc away from the pulsar by the wind. The efficient confinement of electrons was ascribed to a toroidal structure of the magnetic field (perpendicular to the flow velocity), which could be wrapped up by the fast-rotation of the pulsar already in the upstream of the termination shock, and hence the cross-field diffusion of electrons is largely prohibited, or the escape of electrons is inefficient \citep{VanEtten11}. On the other hand, the magnetic field could be highly disturbed by turbulence generated in the pulsar wind, and consequently electrons can escape the plerion and diffuse to large distance away from the pulsar. In the latter scenario, given the true age of the pulsar to be, for instance, 50\,kyr, the escaping electrons could produce a $\sim 100$\,pc--sized nebula with a diffusion coefficient of $10^{28}\rm cm^2s^{-1}$. Therefore, we may avoid assuming an extremely large plerion by invoking unusual conditions such as a very small density for the ambient ISM or a huge kinetic energy for the pulsar wind.

Interestingly, the southern side of the nebula shows intriguing energy-dependent morphology, as revealed by the latest observation of the HESS experiment \citep{HESS19}. The radial profile of the nebula emission at different energy is extracted from a semi-circular region for the southern half nebula. Based on the radial profile, the radial extent of the nebula is measured as the radius from the pulsar at which the flux drops to $1/e$ of the peak value \citep{HESS19}. Such a measurement is crucial to test the transport mechanism of electrons inside the nebula. Furthermore, the energy-dependent behaviour of the nebula's morphology is also related to the pulsar's spin-down history and hence can be used to study the properties of the pulsar such as its true age and the braking index, as will be discussed in this paper. We notice that \citet{HESS19} suggested that the energy-dependent extent of the nebula favours an advection-dominated transport over a diffusion-dominated transport. Their result will be also discussed in detail. 

The rest of this paper is organized as follows: in Section 2, we introduce the basic setup for the injection of electrons. We fit the observed features of the nebula in the pure diffusion scenario, and study the influences of various model parameters on the energy-dependent extent of the nebula in Section 3. In Section 4, we discuss the difference between our results and that obtained in \citet{HESS19}, and the possible contribution from a compact plerion to the observed spectrum. In Section 5, we summarize the results of this work.


\section{Electron injection and  distribution}

\subsection{Injection of electrons from the pulsar}
Let us start with a brief review of the spin-down behaviour of a pulsar, as it determines the injection history of electrons from the pulsar.

The rotational energy of a pulsar is given by
\begin{equation}
W_s=\frac{1}{2}I\Omega^2
\end{equation}
with $I$ being the pulsar's moment of inertia and $\Omega\equiv 2\pi/P$ being its angular velocity. It is generally assumed that $\Omega$ evolves temporally as $\dot{\Omega}=-A\Omega^n$ where $A=
(\dot{P}/P)(P/2\pi)^{n-1}$ is a constant and $n$ is the braking index which is generally $1\lesssim n\lesssim 3$ for some pulsars with reliable measurement on pulsar's spin down \citep{Magalhaes12, Hamil15}. Neglecting the influence of possible glitches or accretion process of the pulsar, the age of the pulsar can be given by
\begin{equation}
t_{\rm age}=\left\{
\begin{array}{ll}
2\tau_c{\rm ln}(\frac{P}{P_0}), \quad n=1\\
\frac{2\tau_c}{n-1}\left[1-\left(\frac{P_0}{P}\right)^{n-1} \right], \quad n\neq 1
\end{array}
\right.
\end{equation}
where $P_0$ is the initial rotation period of the pulsar.

\begin{figure}
\centering
\includegraphics[width=1\columnwidth]{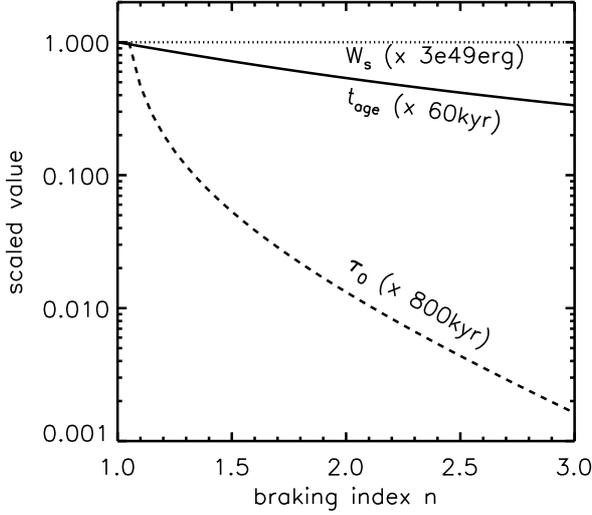}
\caption{Age of the pulsar PSR~J1826-1334 $t_{\rm age}$ (the solid curve), its initial spin-down timescale $\tau_0$ (the dashed curve) and the total released spin-down energy $W_s$ (the dotted curve) as a function of the braking index $n$. The initial rotation period is assumed to be $P_0=25$\,ms here. Note that $\tau_0$ is defined for $n\neq 1$ so its value is normalized at $n=1.05$ while other two quantities are normalized at $n=1$.}\label{fig:braking_evol}
\end{figure}

The spin-down luminosity, which is defined as the rate of the rotational energy being dissipated, can be described by
\begin{equation}\label{eq:Levol}
L_{s,t}(t)\equiv -I\Omega\dot{\Omega}=\left\{
\begin{array}{ll}
L_{s,0}e^{-t/\tau_c}, \quad n=1\\
\frac{L_{s,0}}{\left(1+t/\tau_0 \right)^{\frac{n+1}{n-1}}}, \quad n\neq 1
\end{array}
\right.
\end{equation}
where $\tau_0\equiv P_0/(n-1)\dot{P}_0=2\tau_c/(n-1)-t_{\rm age}$ is the initial spin-down timescale of the pulsar.
The total released spin-down energy up to date is
\begin{equation}
W_s=L_s\tau_c\left[\left(\frac{P}{P_0}\right)^2-1 \right].
\end{equation}
independent of the braking index once the initial period of the pulsar is provided. 
The braking index $n$ is important to the electron injection history. We show the pulsar's age, initial spin-down timescale and total spin-down energy released up to date as a function of the braking index in Fig.~\ref{fig:braking_evol}.

A fraction of the spin-down energy $\eta_e(\leq 1)$ can be converted to the energy of electrons at the termination shock. We then assume that electrons are injected at rate of 
\begin{equation}\label{eq:inj_spec}
Q_{\rm inj}(E_e,t)\equiv\frac{dN}{dE_edt}=Q_0(t)E_e^{-p}, \quad  E_{e,\rm min}\leq E_e \leq E_{e,\rm max}
\end{equation}
with $E_{e,\rm min}$ and $E_{e,\rm max}$ being the minimum energy and maximum energy in the injection spectrum respectively. $p$ is the spectral index. The normalization factor $Q_0(t)$ can be found by $\int E_eQ_{\rm inj}(E_e,t)dE_e=\eta_e L_{s,t}(t)$. 

\subsection{Electron distribution in the pure diffusion scenario}
\begin{figure}
\centering
\includegraphics[width=1\columnwidth]{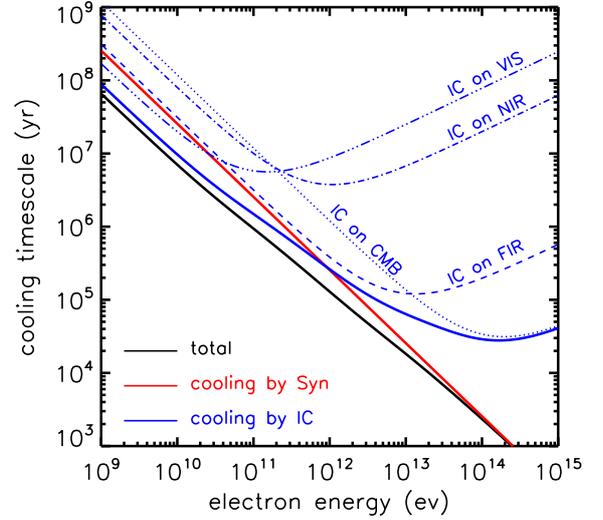}
\caption{Cooling timescale of electrons as a function of electron energy. The red and blue solid curves show the cooling timescale due to synchrotron radiation and IC radiation respectively, while the black solid curve represents the cooling timescale including both processes. We also present the IC cooling timesales due to different background radiation field separately, with the blue dotted curve for CMB, the blue dashed curve for FIR, the blue dot-dashed curve for NIR, and the blue dot-dot-dot-dahsed curve for VIS. The magnetic field is assumed to be $7\mu$G as implied by the X-ray observation.}\label{fig:cooling}
\end{figure}

In this section, we look into the scenario that the particle transport is dominated by diffusion. The diffusion coefficient in the entire TeV emission region is assumed to be homogeneous with the form $D(E_e)=D_0(E_e/1\rm TeV)^\delta$, where $D_0$ and $\delta$ are treated as free parameters. Since we focus on the southern part of the nebula, we only consider particle diffusion in a semi-spherical region assuming isotropic diffusion. Given that the north-south asymmetry of the nebula is due to the crushing by an asymmetric reverse shock interaction \citep{Gaensler03, HESS06_PWN}, such a one-sided diffusion is possible as the magnetic field in the northern part of the nebula may become tangential and the strength is enhanced after the nebula is significantly compressed by the reverse shock \citep[e.g.][]{Reynolds84, Blondin01, Bucciantini03, Vorster13}. The diffusion of electrons to the northern side is then prohibited and hence they would preferably diffuse into the southern side of the nebula. During their propagation, the injected electrons will suffer radiative loss of energy, mainly via the synchrotron radiation and the IC radiation. The energy loss rate is given by
\begin{equation}\label{eq:cooling}
\frac{dE_e}{dt}=-\frac{4}{3}\sigma_Tc\left(\frac{E_e}{m_ec^2}\right)^2\left[U_B+U_{\rm ph}/\left(1+4\frac{E_e\epsilon_0}{m_e^2c^4}\right)^{3/2}\right]
\end{equation}
where $\sigma_T$ is the Thomson cross section, $m_e$ is the electron mass and $c$ is the speed of light. $U_B=B^2/8\pi$ is the magnetic field energy density and $U_{\rm ph}$ is the radiation field energy density. $\epsilon_0=2.82kT$ is the typical photon energy of the radiation field given a black body or a grey body radiation field with a temperature $T$ and $k$ is the Boltzmann constant \citep{Moderski05}. X-ray observation on this nebula by Suzaku suggested a magnetic field of $B=7\mu$G \citep{Uchiyama09} so we assume a constant and homogeneous magnetic field of this strength for the entire region. Following \citet{HESS19}, we consider four black body or grey body components for the radiation field in the location of HESS~J1825-137: the CMB radiation field ($T=2.73\,$K and $U=0.25\, \rm eV cm^{-3}$); a far-infrared radiation field ($T=40$\,K and $U=1\,\rm eV cm^{-3}$); a near-infrared radiation field ($T=500$\,K, $U=0.4\,\rm eVcm^{-3}$); a visible light radiation field (VIS, $T=2500\,$K, $U=1.9\,\rm eVcm^{-3}$). {For reference of the later discussion, we show the electron cooling timescale $t_c$ as a function of energy under this setup in Fig.~\ref{fig:cooling}. }

The present-day ($t=t_{\rm age}$) density of electrons with energy $E_e$ at a radius $r$ away from the pulsar can be calculated by
\begin{equation}\label{eq:diff_sol}
N(E_e,r)=\int_0^{t_{\rm age}} \frac{Q_{e,\rm inj}(E_g,t)dt}{(4\pi \lambda(E_e,t))^{3/2}}\exp\left[-\frac{r^2}{4\lambda(E_e,t)}\right]\frac{dE_g}{dE_e}
\end{equation}
with $\lambda(E_e,t)=\int_t^{t_{\rm age}}D(E_e'(t'))dt'$. Here, $E_e'(t')$ represents the trajectory of energy evolution of an electron the energy of which is $E_e$ at present, and $E_g$ is the initial energy of the electron at the generation (injection) time $t$. The relation between $E_e$ and $E_g$ as well as $dE_g/dE_e$ can be found by tracing the energy evolution of the electron via Eq.~(\ref{eq:cooling}) \footnote{if the IC cooling is limited in the Thomson regime and the injection luminosity is constant over time, an analytical expression for Eq.~\ref{eq:diff_sol} can be obtained, as given by \citet{Atoyan95,Aharonian95}.}. We here neglect the proper motion of the pulsar. This is because the distance travelled by TeV-emitting electrons (with energy $\lesssim 10\,$TeV) is about $2\sqrt{Dt_c}\simeq 50(D/{\rm 10^{28}cm^2s^{-1}})^{1/2}(t_c/{20\rm kyr})^{1/2}\,$pc before cooling, while the proper motion leads to  a shift of the pulsar's position by only $\approx 9(v_{\rm p}/440{\rm km~s^{-1}})(t_c/20\rm kyr)\,$pc with $v_{\rm p}\simeq 440\,\rm km~s^{-1}$ being the velocity of the pulsar's proper motion \citep{Pavlov08}, which is significantly smaller than the former one.

\begin{figure*}
\centering
\includegraphics[width=0.48\textwidth]{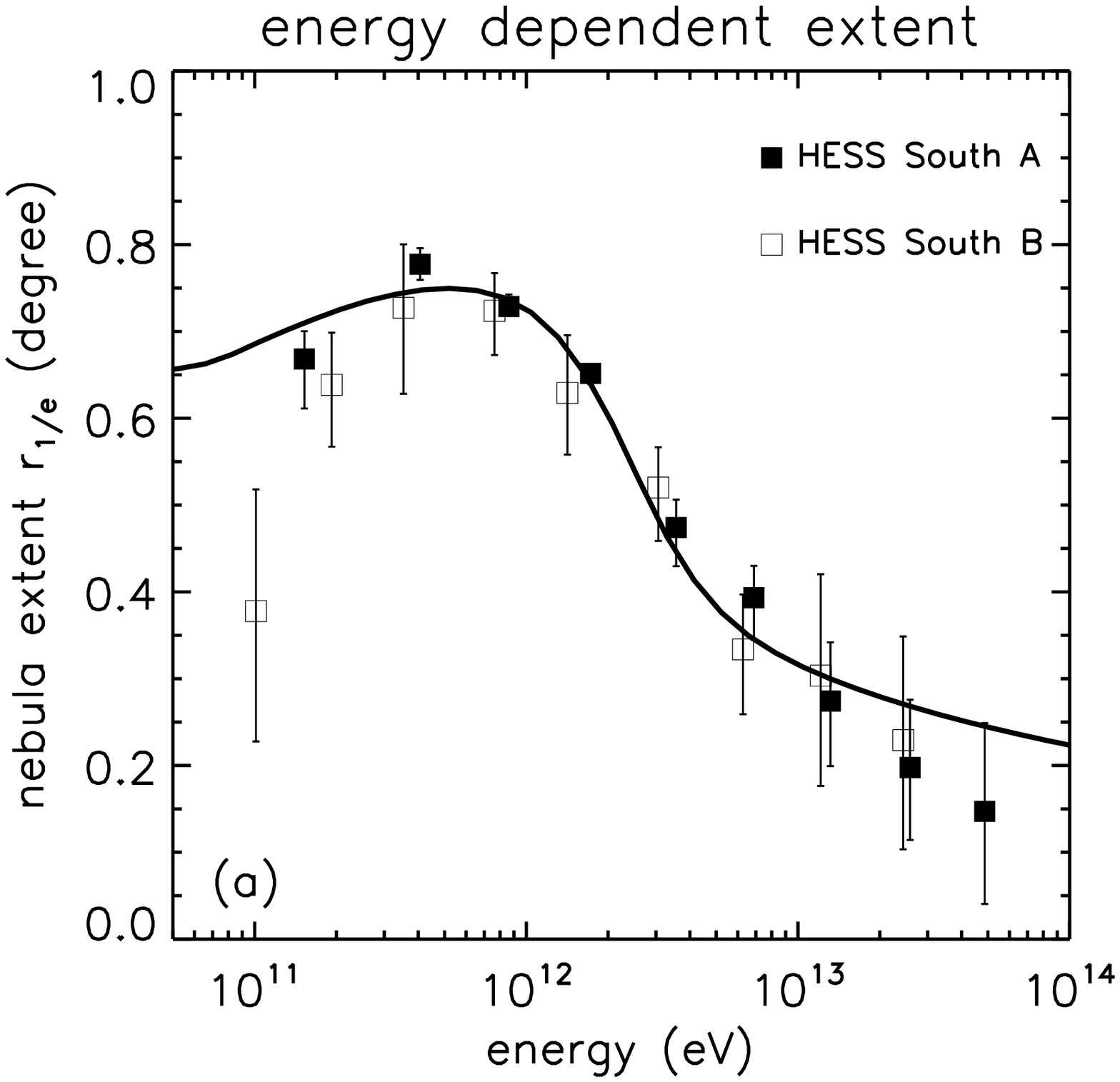}
\includegraphics[width=0.48\textwidth]{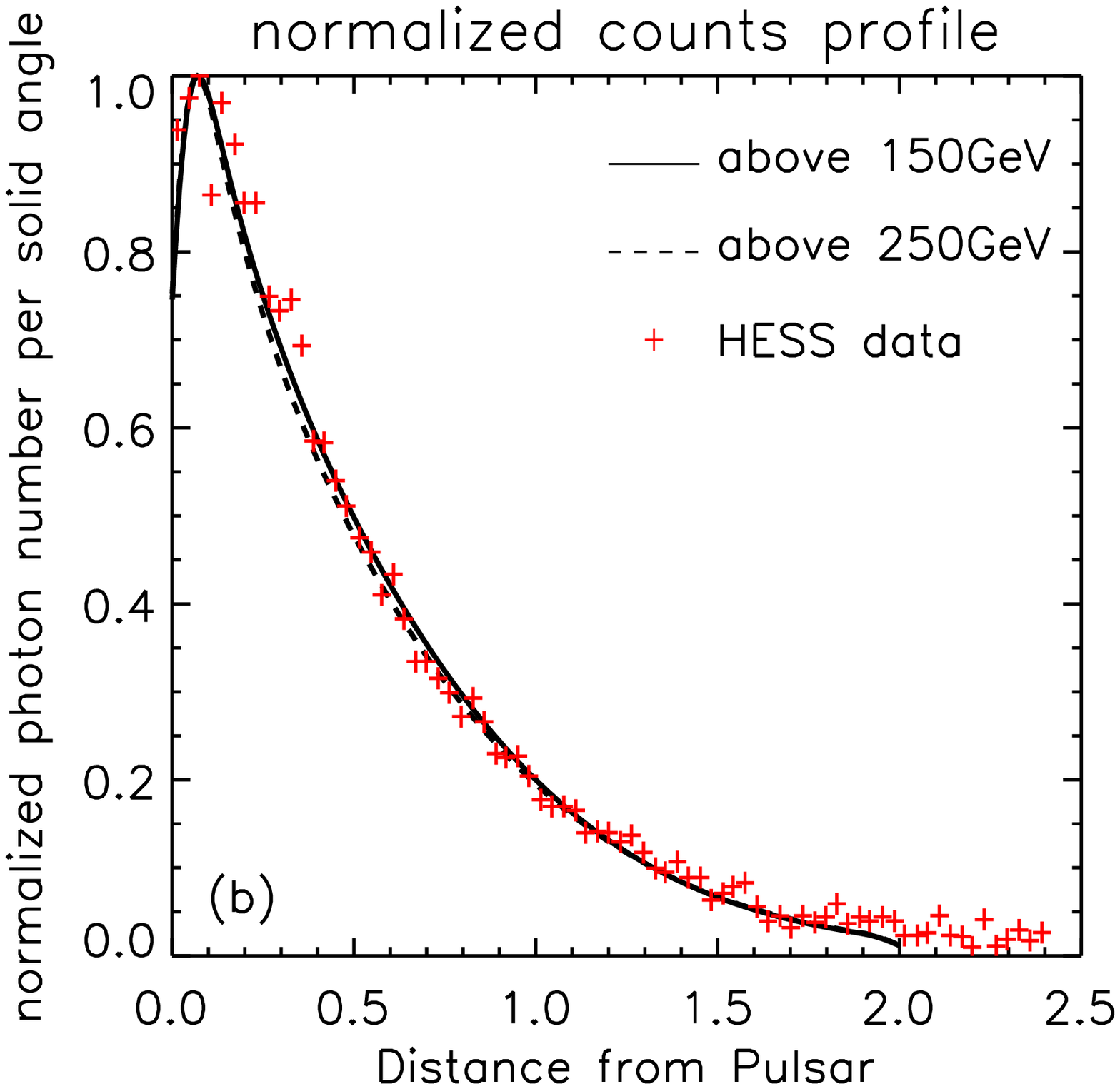}
\includegraphics[width=0.48\textwidth]{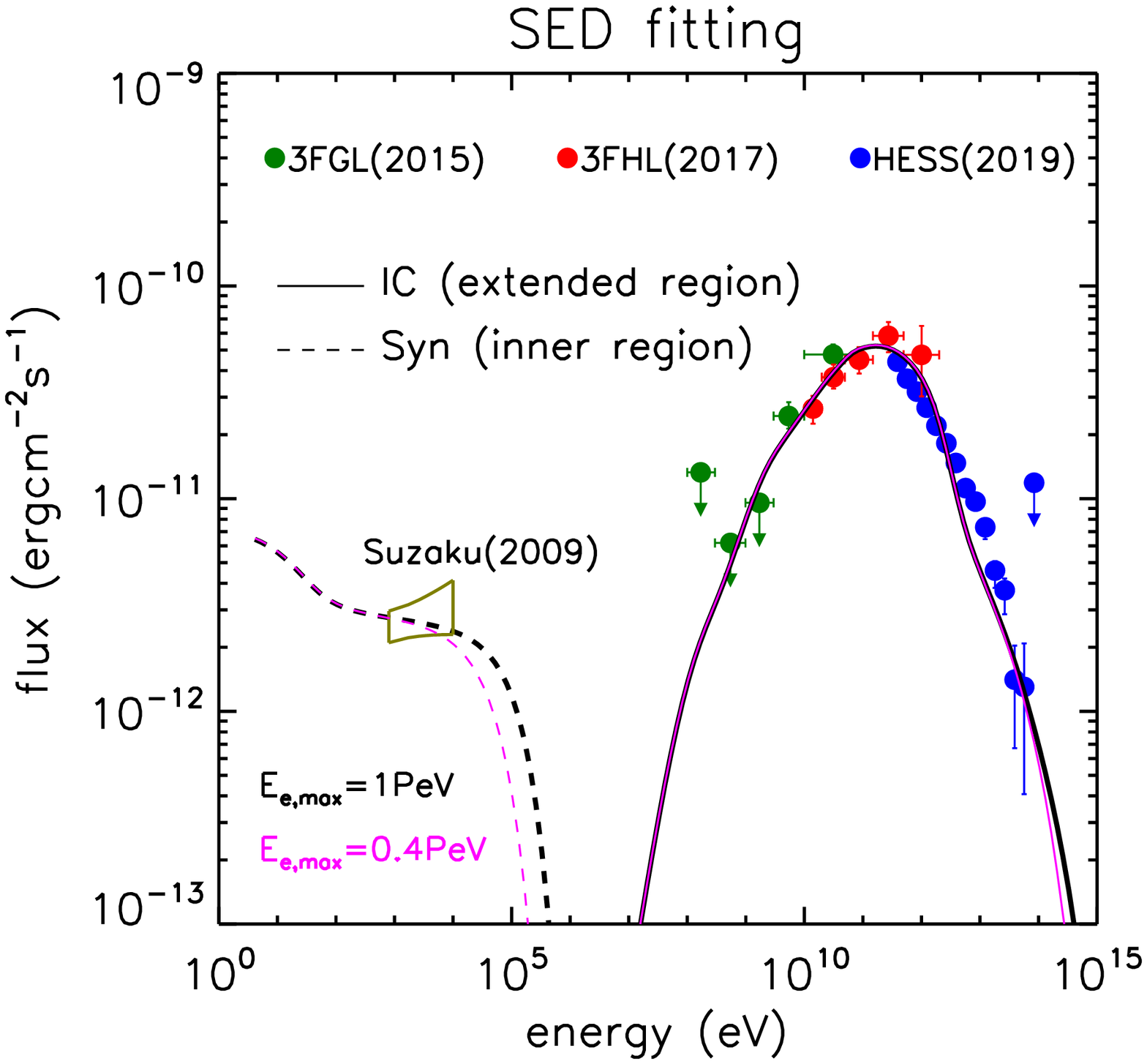}
\includegraphics[width=0.48\textwidth]{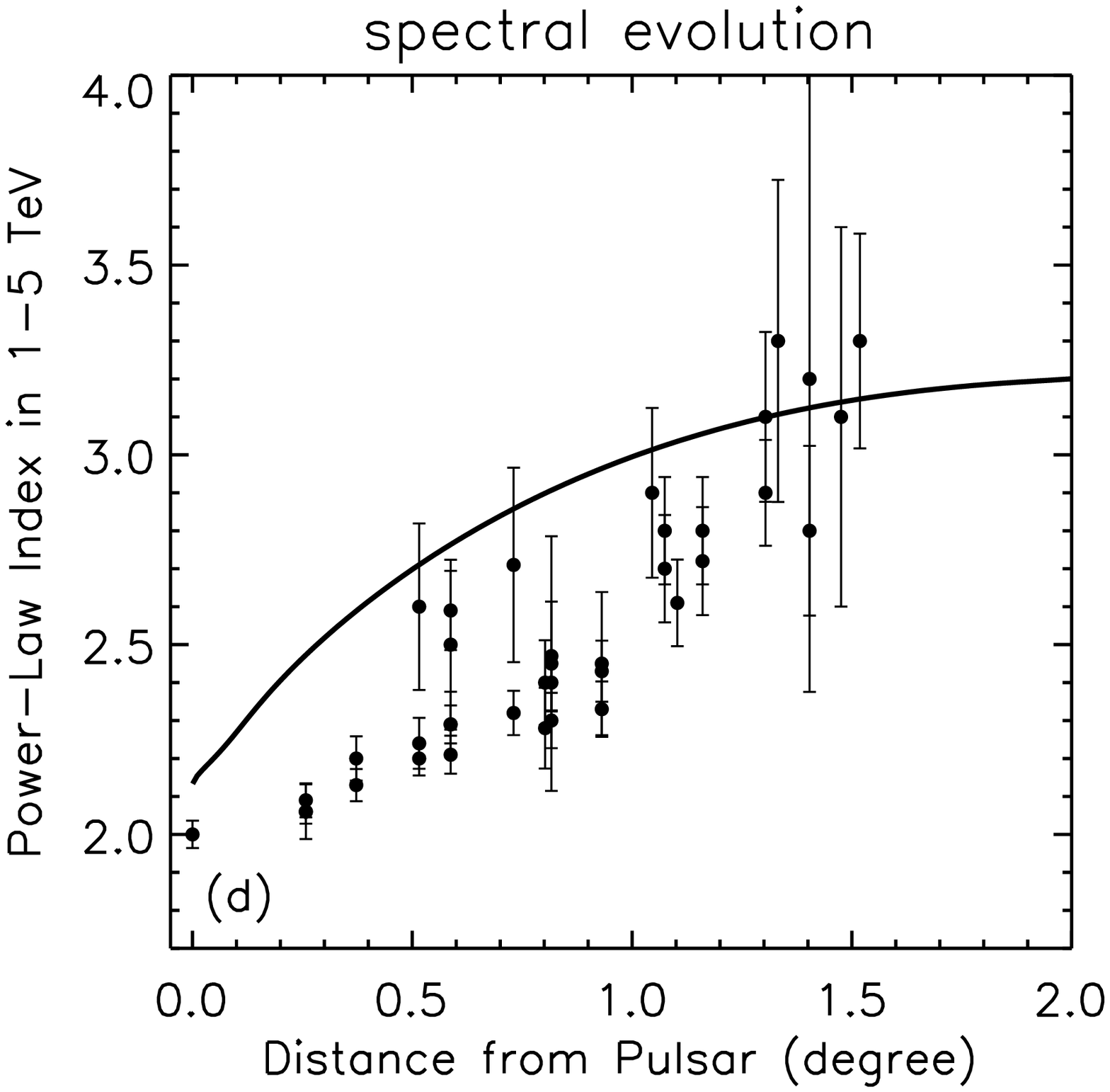}
\caption{{\bf Panel~a (top left):} Predicted energy-dependent extent of the PWN in the benchmark case of the pure diffusion scenario v.s. the measurement of HESS. Open and filled squares are measurements of the extent by HESS using their analysis A and B \citep[see][for details]{HESS19}, respectively;{\bf Panel~b (top right):} Comparison between the predicted number intensity profile of photons above 150\,GeV (solid curve) and the measured counts profile (red crosses); {\bf Panel~c (bottom left):} Fitting to the SED of HESS~J1825-137 in the benchmark case. Filled circles are SED data retrieved from \citet[][green]{Fermi15_3FGL}, \citet[][red]{Fermi17_3FHL} and \citet[][blue]{HESS19}, while the yellow butterfly is the measurement of Suzaku\citep{Uchiyama09} for a region within 15' from the pulsar;  {\bf Panel~d (bottom right):} spectral index in from 1 to 5\,TeV as a function of distance from pulsar. The solid curve is the model prediction while the data points are measurement. Model parameters for the benchmark case: diffusion coefficient $D(E)=10^{28}(E/1{\rm \,TeV})^{0.4}\rm cm^2s^{-1}$, initial rotation period of the pulsar $P_0=25\,$ms, braking index $n=2$, injection spectral index $p=2$, electron conversion efficiency $\eta_e=0.7$.}
\label{fig:fitting}
\end{figure*}

\section{Energy-dependent Morphology of the Neubla}\label{sec:diff}
After obtaining the distribution of electrons, we calculate their IC radiation and integrate the radiation over the line of sight towards an arbitrary position around the pulsar to get the intensity map, following the method detailed in \citet{Liu19}. For simplicity, we assume the angle between the symmetric axis of the semisphere (i.e., the direction in which the nebula extends) and the line of sight of the observer to the pulsar, as denoted by $\phi$, to be $90^\circ$, while we briefly discuss the influence of $\phi$ in the Appendix. To compare with the observation of HESS, the obtained theoretical intensity distribution needs to be convolved with the point spread function (PSF) of HESS. A 2D Gaussian function with a 68\% containment radius of $0.07^\circ$ is adopted \citep[][and private communication with the authors]{HESS19} for the convolution (see Appendix for details). In addition to smoothing the intensity map, the convolution also leads to an offset between the peak in the intensity map and the pulsar location at a distance of $0.1^\circ$ which is comparable to the size of the PSF. This is because photons from the intrinsic peak position, i.e., the pulsar location, spread to the northern side without being compensated by photons from the northern side. This feature is consistent with the observation. 

Averaging over the PSF-convolved intensity map with respect to the azimuthal angle, we obtain the radial profile of the nebula. We then find out the distance to the pulsar (i.e., $r_{1/e}$) where the gamma-ray intensity drops to $1/e$ of the peak value at each energy, following the definition of the nebula extent in \citet{HESS19}. In panel~a of Fig.~\ref{fig:fitting}, we show a satisfactory fitting to the energy-dependent extent of the nebula with diffusion coefficient $D(E)=10^{28}(E/{\rm 1TeV})^{0.4}\rm cm^2s^{-1}$, the initial rotation period of the pulsar $P_0=25\,$ms, the braking index of the pulsar $n=2$, the spectral index of the injection electrons $p=2.4$ and the electron conversion fraction $\eta_e=0.7$. 

We also confront the radial counts profile (counts/arcmin$^2$) of the nebula measured by HESS with the theoretical one in the panel~b of Fig.~\ref{fig:fitting}. The lowest-energy photon considered in the HESS analysis of the counts profile is $<250\,$GeV. Since we do not know the exact value, we show two theoretical counts profiles  with the lowest energy being $150\,$GeV (the solid curve) and $250\,$GeV (the dashed curve) respectively. The difference between these two curves is not significant, because the spectrum below a few hundred GeV is quite hard. Due to lack of the knowledge on the effective area of HESS in the analysis, we normalize the largest value of both the data points and the theoretical profile at unity in order to make them comparable. We can see that the normalized theoretical counts profile is consistent with the data except at large distance ($\gtrsim 2^\circ$ or 140\,pc for a nominal distance of 4\,kpc) where the theoretical profile drops faster than the measured one. Such a deviation is perhaps due to the influence of the background on the measured counts profile at large distance. 

The corresponding multiwavelength flux is exhibited in the panel~c of Fig.~\ref{fig:fitting}. We can see that the measured SED is generally in good agreement with the theoretical one. Note that the spectrum extracting region of the X-ray emission is much smaller than that of the TeV emission. The former focuses on the inner region of the nebula within a distance of 15' to the south of the pulsar. The morphology of the region is close to a sector with a total solid angle of 144~arcmin$^2$ \citep{Uchiyama09}. Thus, to compare with the measured X-ray flux, we integrate the simulated synchrotron radiation intensity over a semi-circular region with a radius of 15' to the south of the pulsar. We further multiply by a factor of 0.45 to correct the difference in the solid angle between the semi-circular region and the sector region where the X-ray spectrum is extracted. {In the SED fitting, $E_{e,\rm min}$ and $E_{e,\rm max}$ are set to 50\,GeV and 1\,PeV respectively. The minimum energy corresponds to the bulk Lorentz factor of the cold pulsar wind (i.e., $=E_{e,\rm min}/m_ec^2$) at the termination shock, which is constrained to be $\sim 10^3-10^6$ for various PWNe \citep{Wilson78, Tanaka11}. The employed $E_{e,\rm min}$ here corresponds to a bulk Lorentz factor of $10^5$ for the pulsar wind, which is in the reasonable range. A smaller $E_{e,\rm min}$ will significantly increase the requirement for the energy budget given $p>2$. Since the obtained value of $\eta_e(=0.7)$ is already close to unity, a smaller $E_{e,\rm min}$ would transcend the energy budget for electrons and hence is not favoured in our model. On the other hand, $E_{e,\rm max}=1\,$PeV is chosen in light of the detection of a $\sim 400$\,TeV gamma-ray photon from the Crab nebula \citep{Asgamma19}. Given $p>2$, the value of $E_{e,\rm max}$ does not affect the energy budget. But it should not be much smaller than $400\,$TeV, since otherwise it would be at odds with the 50\,TeV gamma-ray flux and the 10\,keV X-ray flux, as is illustrated with the magenta curves in panel~c.} 

Lastly, we look into the spatial evolution of the TeV spectrum. In \citet{HESS19}, this is evaluated by the slope of the spectrum in the range of $1-5$\,TeV  as a function of the distance from the pulsar. We therefore fit the theoretical spectrum from 1 to 5\,TeV with a power-law function at each distance from the pulsar, and compare the obtained spectral index to the measurement (see panel~d of Fig.~\ref{fig:fitting}). The theoretical curve roughly reproduces the trend of the spectral softening with increasing distance from the pulsar, which is due to the cooling of electrons. However, it predicts a softer spectrum in $1-5$\,TeV range than the observation especially at small distance. We speculate that such a discrepancy may be alleviated by considering the emission from the central plerion. The electron transport inside the plerion is probably dominated by energy-independent advection so the resulting TeV spectrum at small distance can be harder than that in the diffusion scenario. Such a possibility will be further discussed in Section~\ref{sec:discussion}.

We have shown, in general, that it is feasible to reproduce the energy-dependent extent (as well as other observed features) of the TeV nebula HESS~J1825-137 by a simple diffusion model. In this scenario, we provide an alternative explanation for the unusually large spatial extent of HESS~J1825-137, in addition to considering it as an extremely expanded plerion.

\subsection{Influence of various parameters}\label{sec:inf_pd}
The energy-dependent extent is important to study the particle transport mechanism within the nebula. In this subsection we study the influence of various parameters on the energy-dependent extent of the nebula in order to better understand the result. As we will see, the energy-dependent extent is sensitive not only to the diffusion coefficient, but also to the electron injection history or the spin-down history of the pulsar. Hence, the spin-down behaviour of the pulsar can be studied in turn through the measurement of the energy-dependent morphology of the related nebula. For convenience of comparison, we denote the the case shown in Fig.~\ref{fig:fitting} by the benchmark case.

\subsubsection{Diffusion coefficient}\label{sec:DC}
\begin{figure}
\includegraphics[width=1\columnwidth]{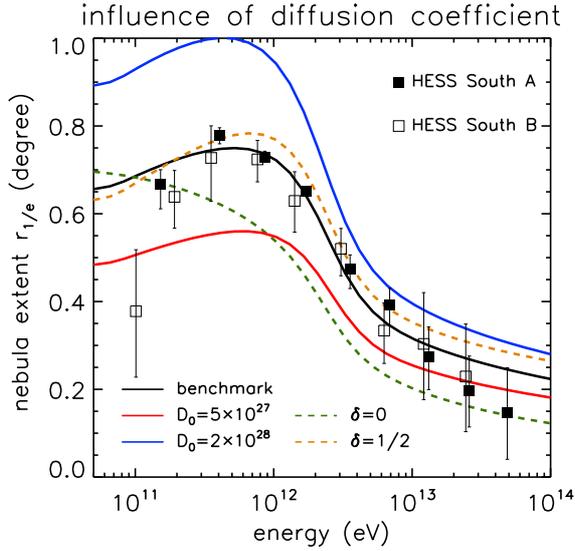}
\caption{Influence of diffusion coefficient on the predicted nebula extent. The black curve is the result in the benchmark case which is the same as the one shown in the upper panel of Fig.~\ref{fig:fitting}. Comparison cases for different $D_0$ are shown with red and blue solid curves, while that for different $\delta$ are shown with green and orange dashed curves. See Section~\ref{sec:DC} for details.}\label{fig:DC}
\end{figure}

The diffusion coefficient has direct impacts on the energy-dependent extension in terms of both $D_0$ and $\delta$.
From Eq.~(\ref{eq:diff_sol}) we can know that $r_{1/e}$ is proportional to $\sqrt{D_0}$ at any energy. This is demonstrated in Fig.~\ref{fig:DC} as we can see that for $D_0=2\times 10^{28}\rm cm^2s^{-1}$ (blue solid curve) the angular extent  systematically increases about a factor of 1.4 compared to the benchmark case in which $D_0=10^{28}\rm cm^2s^{-1}$, while for $D_0=5\times 10^{27}\rm cm^2s^{-1}$ (red solid curve) the angular extent decreases by the same factor. Since we normalize the diffusion coefficient at 1\,TeV, $\delta$ influences the extension at low- and high-energy morphology in different way: a larger $\delta$ increases the extent at high energy and decreases the extent at low energy compared to the benchmark case while a smaller $\delta$ changes the extent in the opposite way, as shown with the green ($\delta=0$) and the orange ($\delta=1/2$) dashed curves in Fig.~\ref{fig:DC}. 

We have shown, in general, that it is feasible to reproduce the energy-dependent extent (as well as other observed features) of the TeV nebula HESS~J1825-137 by a simple diffusion model. In this scenario, we also provide an alternative explanation for the unusually large spatial extent of HESS~J1825-137, in addition to considering it as an extremely expanded plerion.

\subsubsection{Initial rotation period $P_0$}\label{sec:P0}
\begin{figure}
\centering
\includegraphics[width=1\columnwidth]{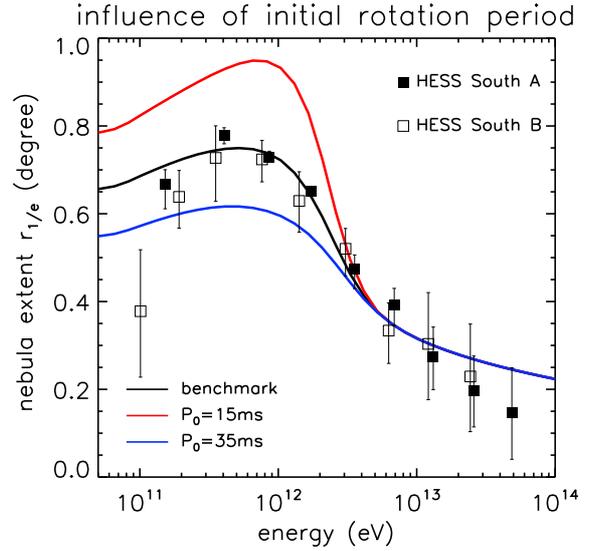}
\caption{Influence of pulsar's initial rotation period $P_0$ on the predicted nebula extent. See Section~\ref{sec:P0} for details.}\label{fig:P0}
\end{figure}

Since the current rotation period $P$ of the pulsar and its time derivative $\dot{P}$ are measured, $P_0$ will influence the age of the pulsar and the initial spin-down luminosity if the braking index is further presumed. A smaller $P_0$ will increase $t_{\rm age}$ and $L_{s,0}$, and vice versa. The morphology at comparatively high energy is not sensitive to the change of $P_0$ since high-energy electrons injected at early time has already cooled. Even if a huge amount of high-energy electrons is injected at early time, they cannot survive to the present day. On the contrary, low-energy electrons diffuse more slowly and cool less efficiently than high-energy electrons. A longer injection history of the pulsar facilitates the transport of low-energy electrons to farther distance from the pulsar within the age of the system and makes the nebula more extended at comparatively low energy. Such an effect can be seen in Fig.~\ref{fig:P0}. Regarding the influence on the gamma-ray spectrum, a smaller $P_0$ will lead to a softer spectrum at the present day because, firstly, the amount of early injected electrons, which has already been well cooled, will be larger, and secondly, the cooling break will appear at lower energy for a larger $t_{\rm age}$.

\subsubsection{braking index}\label{sec:braking}
\begin{figure}
\centering
\includegraphics[width=1\columnwidth]{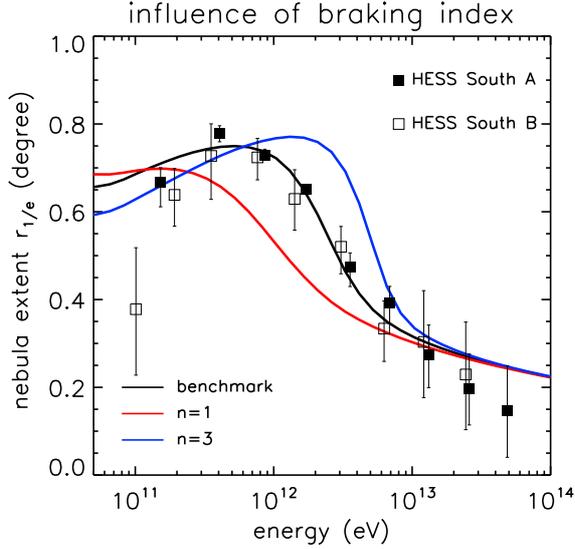}
\caption{Influence of pulsar's braking index $n$ on the predicted nebula extent. See Section~\ref{sec:braking} for details.}\label{fig:braking}
\end{figure}

The braking index $n$ determines the injection profile of electrons and its influence on the energy-dependent morphology is complex. 
For a large braking index, e.g., $n=3$, the age of the pulsar $t_{\rm age}$ is much longer than the initial spin-down timescale $\tau_0$ (see Fig.~\ref{fig:braking_evol}). The spin-down luminosity in this case decreases as $t^{-2}$ except at very early time (Eq.~\ref{eq:Levol}). For a smaller braking index, the spin-down luminosity decreases more quickly with time for $t>\tau_0$. However, $\tau_0$ will become comparable or even larger than the age of the pulsar. As a result, the global decline slope of the spin-down luminosity from the birth of the pulsar to the present day is flatter for a smaller $n$. Note that radiation at smaller (larger) radius mainly arises from electrons injected at later (earlier) time, so a smaller $n$ results in a steeper decrease of the electron density profile and hence leads to a less extended morphology and vice versa. Such a tendency can be seen from Fig.~\ref{fig:braking} except at the low-energy end and the high-energy end. At the low-energy end, the extent of the PWN is mainly determined by the age of the system, so it is larger for a smaller $n$. At the high-energy end, the cooling timescale of electrons is very short (within a few thousand years) so the radiating electrons are just recently injected. In other word, the spatial distribution of high-energy electrons is not sensitive to the long-term temporal behaviour of the spin-down luminosity and hence the curves for all three values of $n$ converge at high-energy end.

\subsubsection{Spectral index}\label{sec:slope}
{We assume a single power-law of index $p$ for the injection spectrum (Eq.~\ref{eq:inj_spec}). The SED shows a peak around $\sim 100\,$GeV which is naturally formed due to cooling of electrons and superposition of electrons injected at different epoch, without invoking a broken power-law function for the injection spectrum. The peak in the SED roughly corresponds to the transition in the slope of the present-day electron spectrum from $<-3$ to $>-3$, so the injection spectral index $p$ can influence the peak energy in the SED. 
We show the present-day electron spectrum ($E_e$ v.s. $E_e^3dN/dE_e$) with decomposing the contributions from different injection epoch in Fig.~\ref{fig:espec} for different $p$. We note that the spectrum shows a hump structure around the cooling break instead of a simple broken power law. This is because the broken power-law spectrum can be formed via synchrotron or IC cooling only when the injection rate of electrons is constant, while the latter decreases with time as the pulsar spins down.  

Electrons that have propagated to larger radius have suffered severer cooling since they were generally injected at earlier time. Thus, for a harder injection spectrum, the amount of electrons at large radius is higher than that in the case with a softer injection spectrum, and subsequently the decline of the present-day electron density along $r$ will be shallower. We therefore anticipate that the nebula extent will be larger (smaller) for a smaller (larger) $p$, as is shown in Fig.~\ref{fig:slope}.

\begin{figure}
\centering
\includegraphics[width=1\columnwidth]{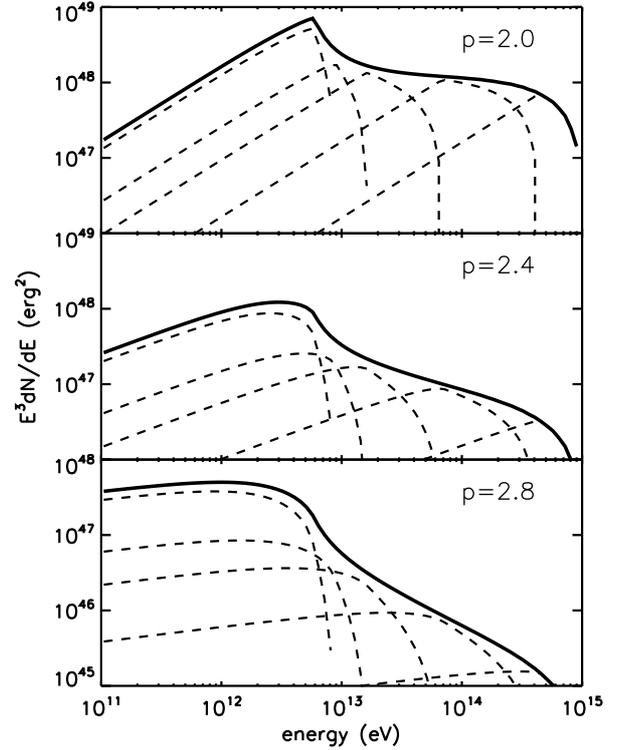}
\caption{Electron spectrum in the nebula at the present day for $p=2.0$ (top), $p=2.4$ (middle, the benchmark case) and $p=2.8$ (bottom), with all other parameters same with the benckmark case. Dashed curves represent the present-day spectrum of electrons injected at different time (in each panel, from top left to bottom right: $t=(0-0.01)t_{\rm age}$, $(0.01-0.1)t_{\rm age}$,$(0.1-0.4)t_{\rm age}$, $(0.4-0.7)t_{\rm age}$, $(0.7-1)t_{\rm age}$), while solid curves are the superposition of the spectrum of injected at different epochs. See Section~\ref{sec:slope} for details.}\label{fig:espec}
\end{figure}

\begin{figure}
\centering
\includegraphics[width=1\columnwidth]{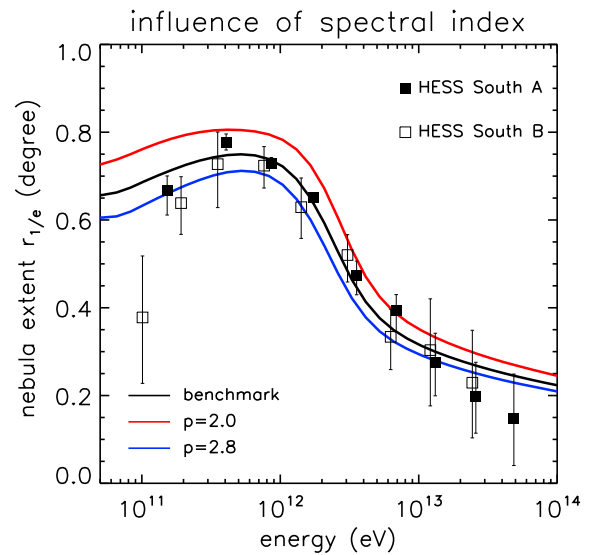}
\caption{Influence of the spectral index of the injection electrons on the predicted nebula extent. See Section~\ref{sec:slope} for details.}\label{fig:slope}
\end{figure}

\section{Discussion}\label{sec:discussion}
\subsection{Comparison with the result in \citealt{HESS19}}
We notice that \citet{HESS19} reach a conclusion that a pure diffusion scenario for particle transport is not favoured by the energy-dependent extent of the nebula, also based on the 1D particle transport model. The discrepancy mainly arises from different ways of modelling the extent of the nebula. In this paper, we firstly calculate the distribution of electrons and obtain the intensity profile by performing the line-of-sight integration of the radiation of the electrons at different radius, and finally find at which radius the intensity drops to $1/e$ of the peak intensity. The obtained extent in this way follows the same definition of the measured one. On the other hand, \citet{HESS19} mainly focus on the energy dependence of the extent, assuming the latter to be determined by the cooling of electrons (i.e., proportional to the distance travelled by electrons within their cooling timescales $r_c$). We'd like to point out that although the cooling of electrons indeed have an impact on the nebula extent, the extent determined in this way does not follow the same definition with the extent given by $r_{1/e}$. Also, the cooling timescales of comparatively low energy electrons (e.g., $\lesssim 10\,$TeV) can be even longer than the pulsar's age (see Fig.~\ref{fig:cooling}). Furthermore, the gamma-ray intensity profile is also influenced by other factors as shown in Section~\ref{sec:inf_pd}. 
Therefore, it may not be appropriate to compare the extent determined by electron cooling with the measured $r_{1/e}$.

\subsection{Possible influence of a central plerion}\label{sec:plerion}
In Section.~\ref{sec:diff}, we showed that the TeV nebula HESS~J1825-137 can be understood as the emission of diffusing electrons which escape from the associated plerion of PSR~J1826-1334, and hence the true size of the plerion is much smaller than that of the extended TeV emission in this scenario. The model, however, predicts a softer spectrum than the observation in $1-5$\,TeV at small distance from the pulsar ($\theta<0.5^\circ$). Such a deviation might be ameliorated by taking into account the emission of the compact plerion. { Simulations show that for a pulsar with spin-down luminosity similar to PSR~J1826--1334 and an ISM density of $0.1-1\, \rm cm^{-3}$, the outer boundary of the plerion could reach $10-20\,$pc at several tens of thousand years \citep{Vorster13_simulation}, corresponding to $\lesssim 0.3^\circ$ at a nominal distance of 4\,kpc. The particle transport within the plerion may probably be dominated by energy-independent advection \citep{KC84, Porth14}. Some previous literature also suggest energy-independent diffusion of particles inside some young plerions \citep{Tang12, Porth16}. {From theoretical point of view, diffusion with flat dependence can indeed occur in a collisionless environment \citep{Yan08}.}} In these cases, we can generally expect the electron spectrum inside the plerion to be harder than that in the pure diffusion scenario considered in this work, as in the latter scenario the energy-dependent diffusion would soften the spectrum by $E_e^\delta$. 
As a result, the gamma-ray spectrum at small radius would be hardened by $E_\gamma^{\delta/2}$.
On the other hand, electrons are injected into the ambient medium from the surface of the plerion instead of from a point at $r=0$ as considered in the pure diffusion scenario. Therefore, the time needed to travel to certain radius $r$ from the pulsar would be shorter, and consequently the spectrum of electrons outside the plerion would also be systematically hardened compared to that in the pure diffusion scenario. An appropriate modelling of the energy-dependent morphology of the nebula in the case of taking into account a central plerion requires a sophisticated treatment to the electron distribution incorporating both advection and diffusion \citep[e.g.][]{VanEtten11, Ishizaki18}, as well as to the transition of these two transport mechanisms on the surface of the plerion. The velocity profile of the advecting flow inside the plerion is also needed in order to get the distribution of electrons within the plerion. Such a kind of study is beyond the scope of this work and we leave it to the future study.

\subsection{Is HESS~J1825-137 a TeV halo?}
\citet{HAWC17_Geminga} reported discovery of the diffuse emission at multi-TeV energy around the Geminga pulsar and PSR~B0656+14, extending out to at least a distance of 30\,pc around the pulsars. {Such a phenomenon is also called ``TeV halo'' \citep[e.g.][]{Hooper18}, which is firstly predicted by \citet{Aharonian04}.}
The surface brightness profiles of the TeV halos are consistent with the diffusion-dominated transport of electrons escaping the PWNe\citep{HAWC17_Geminga, Lopez18, Tang19, DiMauro19, Liu19_prl}. If HESS~J1825 is indeed produced by escaping electrons as the scenario proposed in this paper, it then belongs to the TeV halo. However, it is worth noting that  PSR~J1826-1334 which powers HESS~J1825-137 is much younger than Geminga and PSR~B0656+14. The characteristic ages of the latter two are, respectively, 342\,kyr and 110\,kyr. The surrounding environment of these two pulsars are basically ISM, because the pulsars have already left the related SNRs in such a long time due to their proper motions, or, even if not, the SNRs themselves are already too old to be energetically important. By contrast, PSR~1826-1334 may still reside well inside the related SNR, where the magnetic field and the turbulence could be stronger than those in the ordinary ISM. {This may explain why the diffusion coefficient obtained in our benchmark case, i.e., $D(E_e)=10^{28}(E_e/{1\rm TeV})^{0.4}\,\rm cm^2s^{-1}$, is about one order of magnitude smaller than the standard ISM diffusion coefficient inferred from the measurement of the secondary-to-primary ratio in the local CR spectrum \citep[e.g.][]{Aguilar16}. On the other hand, we could not rule out the possibility that the surrounding medium of the pulsar is largely ISM instead of interior of a SNR. In this scenario the low diffusion coefficient in the nebula may arise from the streaming instability driven by CRs themselves, as a result of an enhanced CR flux around the source \citep{Yan12}.}

{Note that the best-fit diffusion coefficients in the TeV halos of Geminga and PSR~B0656+14, in the context of isotropic particle diffusion, are found to be only $4.5\times 10^{27}(E_e/100{\rm TeV})^{1/3}\,\rm cm^2s^{-1}$ or $\sim 10^{27}\, \rm cm^2s^{-1}$ at 1\,TeV \citep{HAWC17_Geminga}. This value is even one order of magnitude smaller than the diffusion coefficient of HESS~J1825-137 in our model. \citet{Fang19} indicate that such low diffusion coefficients cannot arise from the streaming instability, but might be driven by the SNRs that the pulsars are inhabiting. If this is true, it would raise an intriguing question that how could those old SNRs with ages $\sim 100\,$kyr generate much stronger turbulences than the SNR related to HESS~J1825-137 with an intermediate age of $\sim 10\,$kyr. Alternatively, we may resort to other possibilities for the low diffusion coefficients of the TeV halos such as certain small scale ($\sim 1$\,pc) turbulence-driving mechanism \citep{Lopez18} or the perpendicular diffusion {in the presence of a sub-Alfv{\'e}nic turbulence} around the pulsars \citep{Liu19_prl}. In any case, HESS~J1825-137 seems probably not a typical TeV halo as those around old pulsars, even though the particle transport in the nebula may be also dominated by diffusion.} Recently, \citet{Giacinti19} compare the energy density of TeV-emitting electrons $\epsilon_e$ and the typical ISM energy density $\epsilon_{\rm ISM}$ of the two TeV halos and TeV PWNe (plerions), and found that $\epsilon_e\ll \epsilon_{\rm ISM}$ for the two TeV halos while $\epsilon_e\gtrsim \epsilon_{\rm ISM}$ for more most of TeV PWNe. Interestingly, they found that HESS~J1825-137 probably locates in the mixed regime between these two situations, which is supportive to our speculation in Section~\ref{sec:plerion}.

\section{Conclusion}
To conclude, we have studied the energy-dependent morphology of the TeV nebula HESS~J1825-137 in a pure diffusion scenario for the transport of electrons within the nebula. The observed features of the nebula can be generally reproduced in this scenario, through ascribing the emission to the escaping electrons from a compact plerion. It provides an alternative explanation for the large extent of the nebula (i.e., $\sim100$\,pc), and may avoid invoking some unusual parameters, e.g., a very low ambient medium density or a huge kinetic energy of the  pulsar wind as when considering the entire nebula to be an extremely expanded plerion. In the pure diffusion scenario, although the model predicts a softer spectrum in $1-5\,$TeV range than the observation, it may be ameliorated by taking into account an additional contribution to the TeV flux by the central plerion. {In this work, we have also demonstrated that the energy-dependent extent of a nebula is sensitive not only to the diffusion coefficient, but also to the spin-down history of the related pulsar, so the latter could be in turn studied through measuring the energy-dependent morphology of the nebula. }

\section*{Acknowledgements}
We thank the referee for constructive suggestions. We also thank Alison Mitchell and Gwenael Giacinti for helpful discussions, and Felix Aharonian for useful comments.

\appendix
\section{Convolving the intensity map with the PSF of HESS}
To compare the theoretical extent of the neubla $r_{1/e}$ and the measured value by HESS, we need to convolve the obtained intensity map with the PSF of HESS. The latter can be depicted by a two-dimensional Gaussian function, i.e.,
\begin{equation}
f_{\rm PSF}(x,y)=\frac{1}{2\pi}\exp\left(-\frac{x^2+y^2}{2\sigma_{\rm PSF}^2}\right)
\end{equation}
where $x$ and $y$ is a coordinate system on the plane of sky, and $\sigma_{\rm PSF}$ is the size of the PSF of HESS, where we adopt the 68\% containment radius $\simeq 1.51\sigma_{\rm PSF}=0.07^\circ$. Denoting the theoretical intensity map by $I_{\rm theo}(x,y)$, the PSF-convolved intensity map $I_{\rm PSF}(x,y)$ can be calculated by
\begin{equation}
I_{\rm PSF}(x,y)=\int_{x_0}\int_{y_0} I_{\rm theo}(x_0, y_0)f_{\rm PSF}(x-x_0, y-y_0) dx_0dy_0.
\end{equation} 
For reference, we show the original theoretical intensity map and the PSF-convolved theoretical intensity map of the benchmark case in Fig.~\ref{fig:intensity_map}.

\begin{figure*}
\centering
\includegraphics[width=0.48\textwidth]{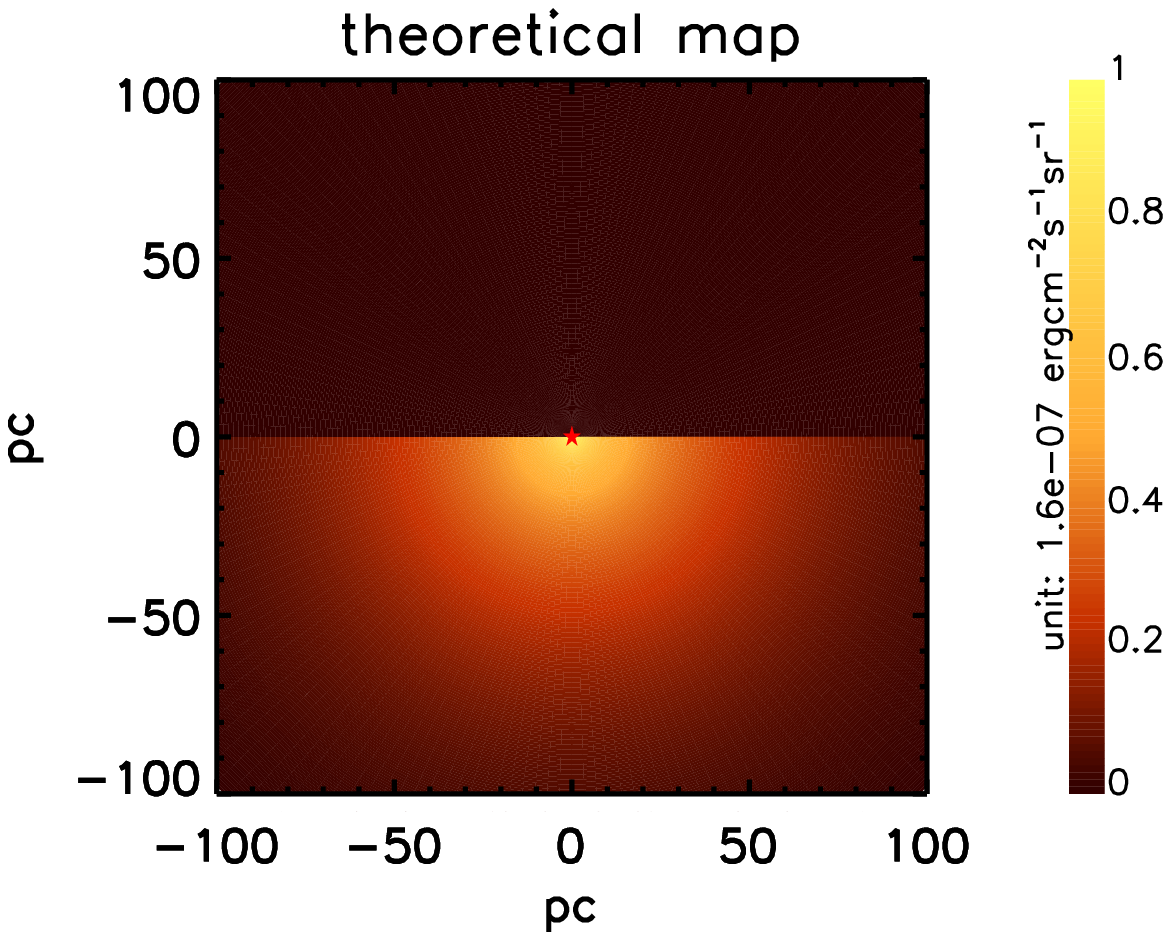}
\includegraphics[width=0.48\textwidth]{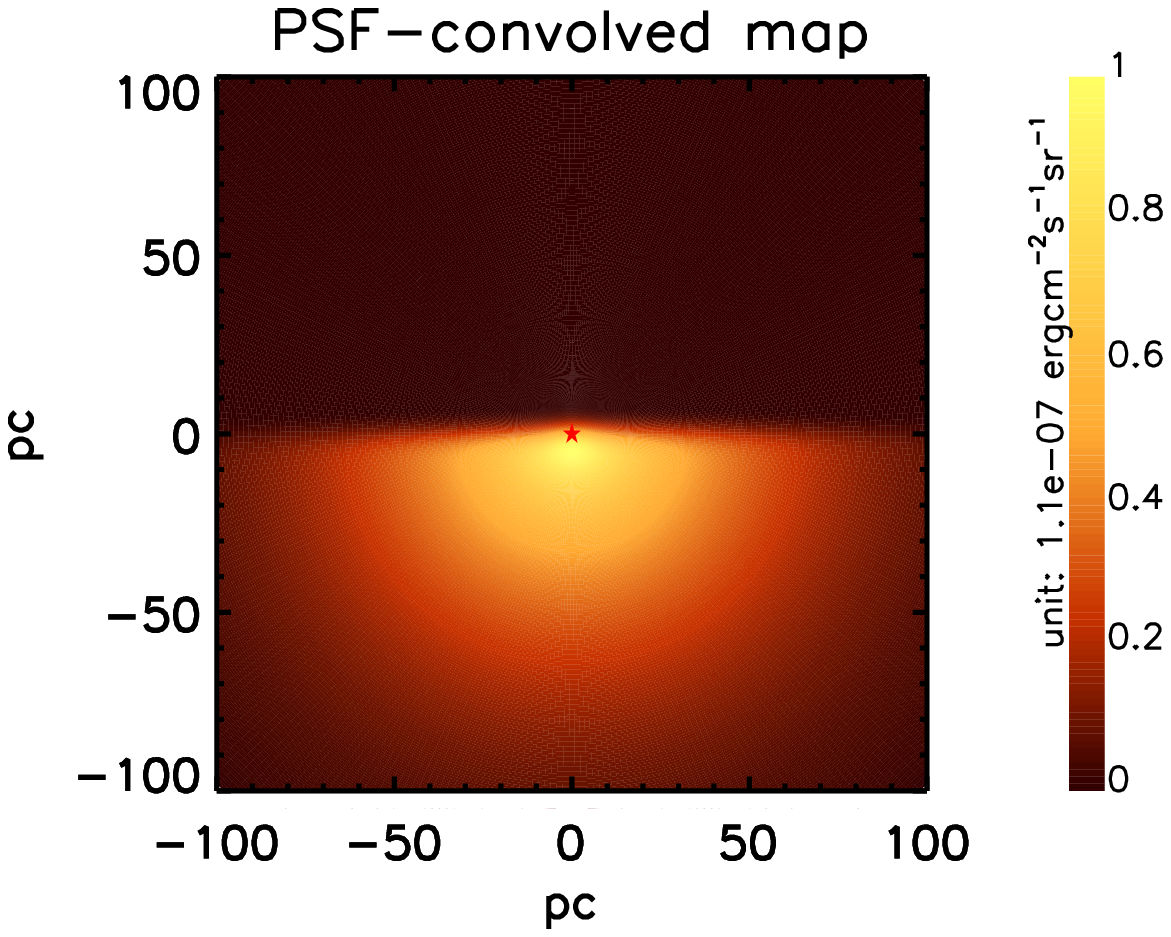}
\caption{Intensity map of 1\,TeV emission in the benchmark case. The left panel shows that of the original intensity map while the right panel shows the PSF-convolved intensity map.}\label{fig:intensity_map}
\end{figure*}

\section{Influence of the viewing angle}
{In the main text of this paper, we focus on the case of a viewing angle of $\phi=90^\circ$ with respect to the symmetric axis of the semispherical nebula for simplicity. The viewing angle can influence the projected morphology of the nebula and hence the expected energy dependent extent. We here present results with different values of $\phi$, noting that the result of viewing angle $\phi>90^\circ$ is the same with that of the viewing angle $(180^\circ-\phi)$ as long as the distance of the pulsar to Earth is much longer than the size of the nebula. All the other parameters are the same with those in the benchmark case. In Fig.~\ref{fig:morphology_compare}, we show the morphology of the nebula with smaller $\phi$ (i.e., $\phi=60^\circ,70^\circ, 80^\circ$). We can see clearly that $\phi<70^\circ$ is not favoured because the projected nebula extends too much to the north of the nebula. The expected count profile and the energy-dependent extent with $\phi=70^\circ$ and $\phi=80^\circ$ are, however, inconsistent with observation with the benchmark parameters, as is shown in Fig.~\ref{fig:extent_compare}. 
Of course, we can adjust other model parameters in the case with $\phi=70^\circ$ and $\phi=80^\circ$ to make a better reproduction of the measured features of the nebula. For instance, in the case of $\phi=80^\circ$, by simply adopting a bit larger initial rotation period for the pulsar, i.e., $P_0=30\,$ms, the predicted counts profile and the energy-dependent morphology can fit the observation better, as shown with the black dashed curves in Fig.~\ref{fig:extent_compare}. We refrain from expanding the discussion on the results with $\phi<90^\circ$, since we cannot constrain the value $\phi$ well without a precise measurement on the northern morphology of the nebula. Besides, another extended TeV source HESS~J1826-130 lies in close proximity to the north of HESS~J1825-137 \citep{HESS19}. This unidentified source may significantly contaminate the emission from on the northern nebula of HESS~J1825-137.}

\begin{figure*}
\centering
\includegraphics[width=0.325\textwidth]{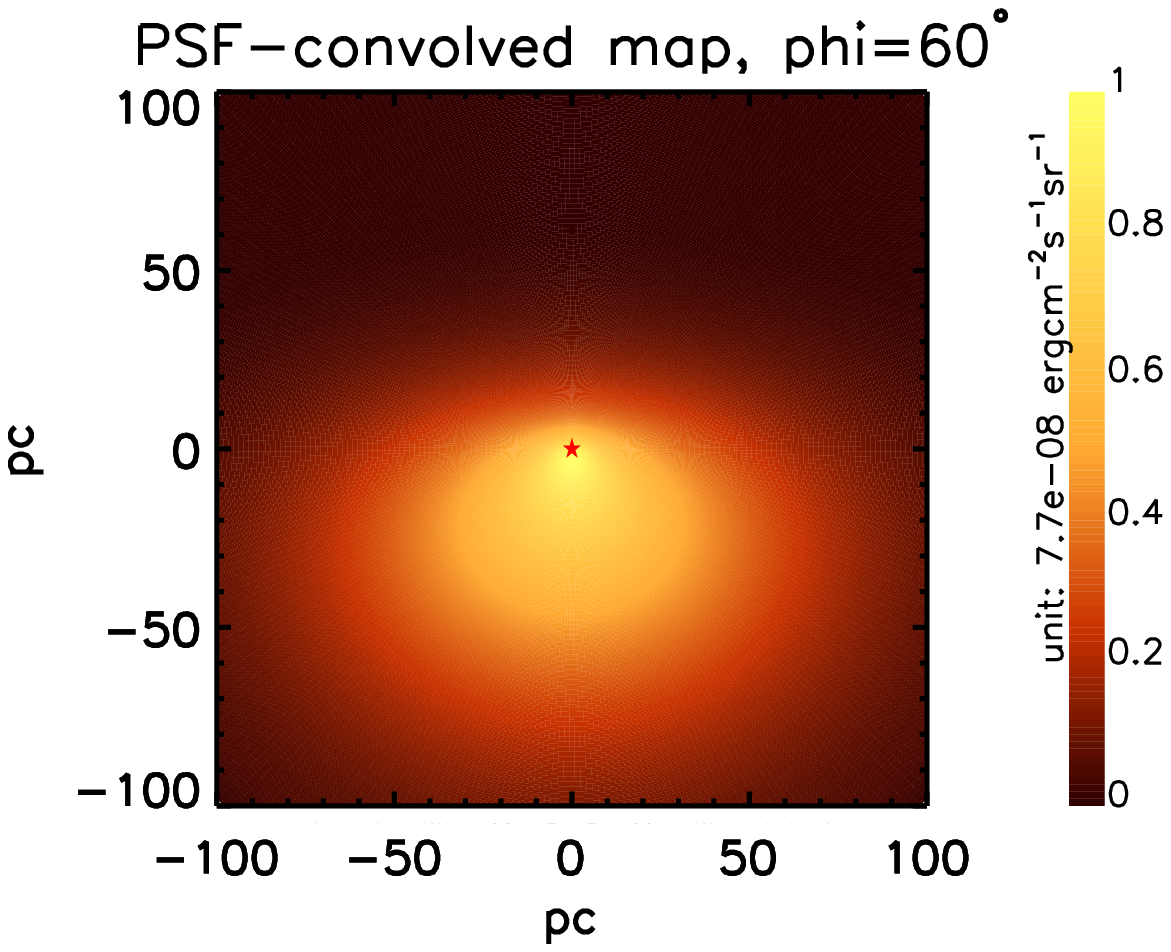}
\includegraphics[width=0.325\textwidth]{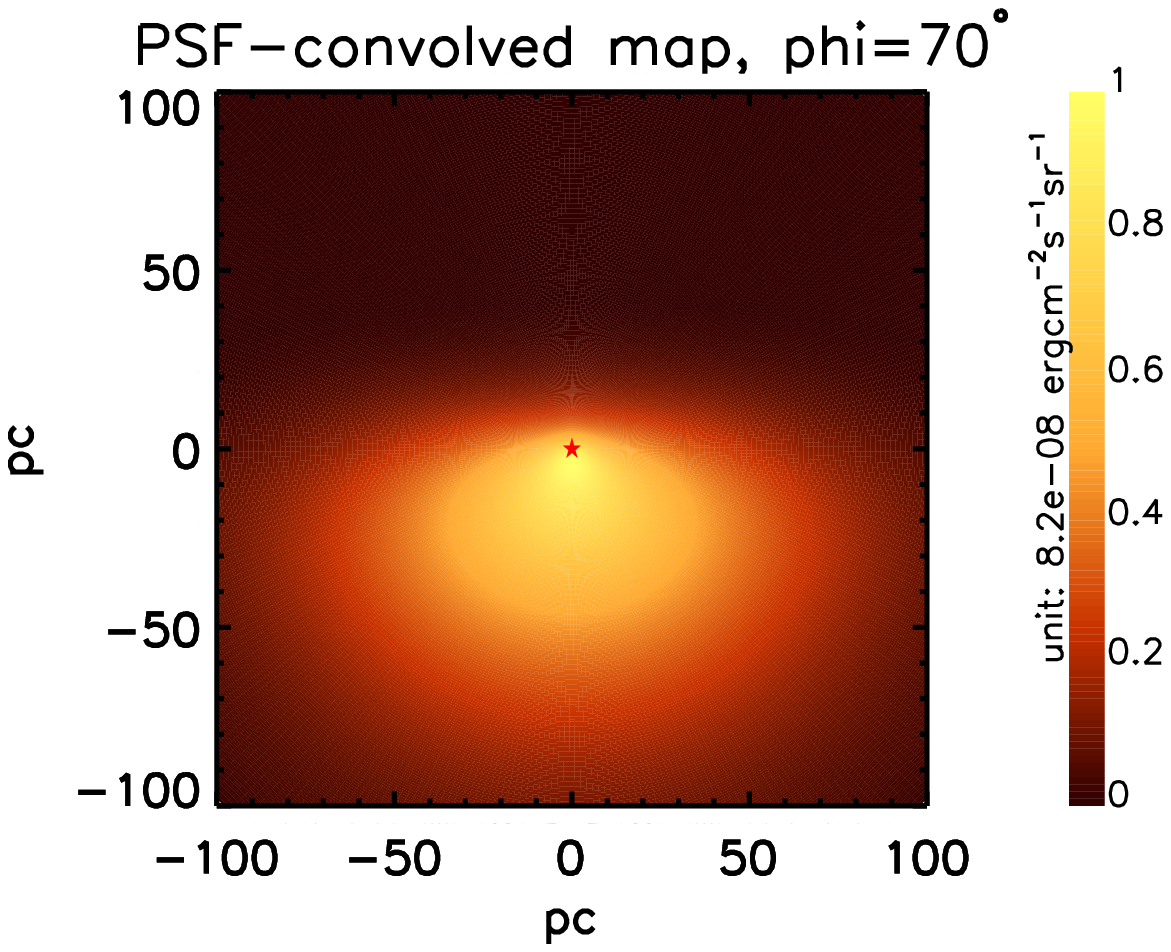}
\includegraphics[width=0.325\textwidth]{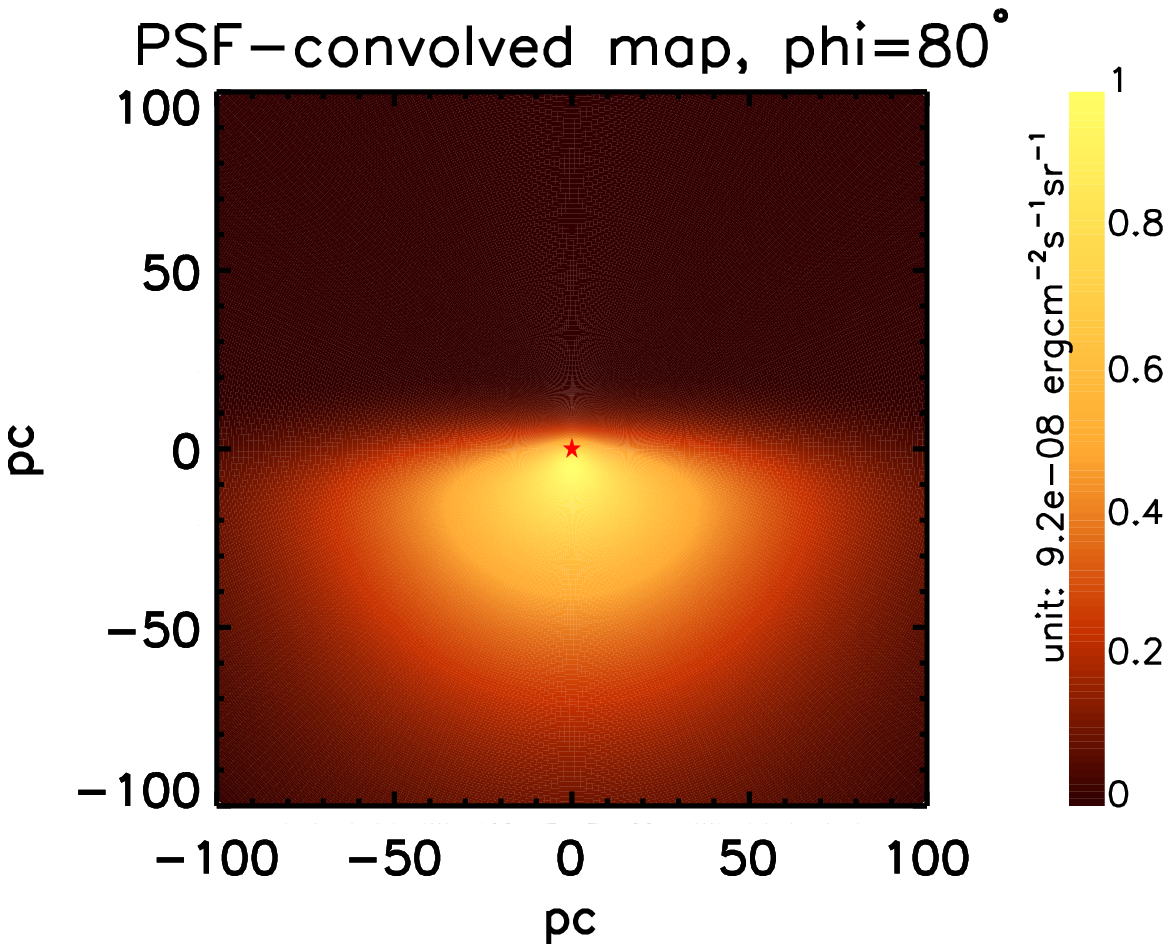}
\caption{PSF-convolved intensity map of 1\,TeV emission under different viewing angle $\phi$. Other parameters follow those in the benchmark case for solid curves.}\label{fig:morphology_compare}
\end{figure*}

\begin{figure*}
\centering
\includegraphics[width=0.48\textwidth]{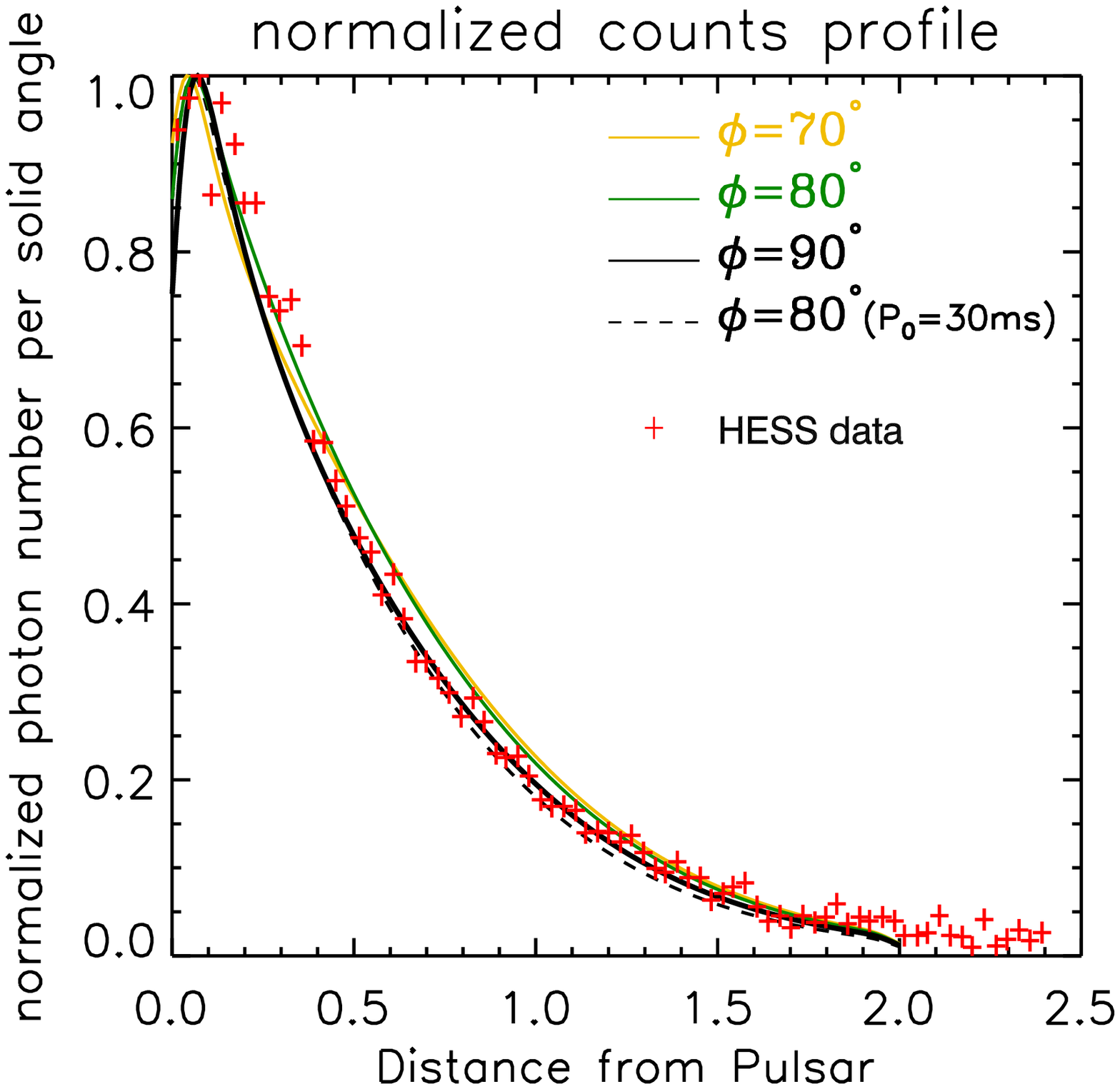}
\includegraphics[width=0.48\textwidth]{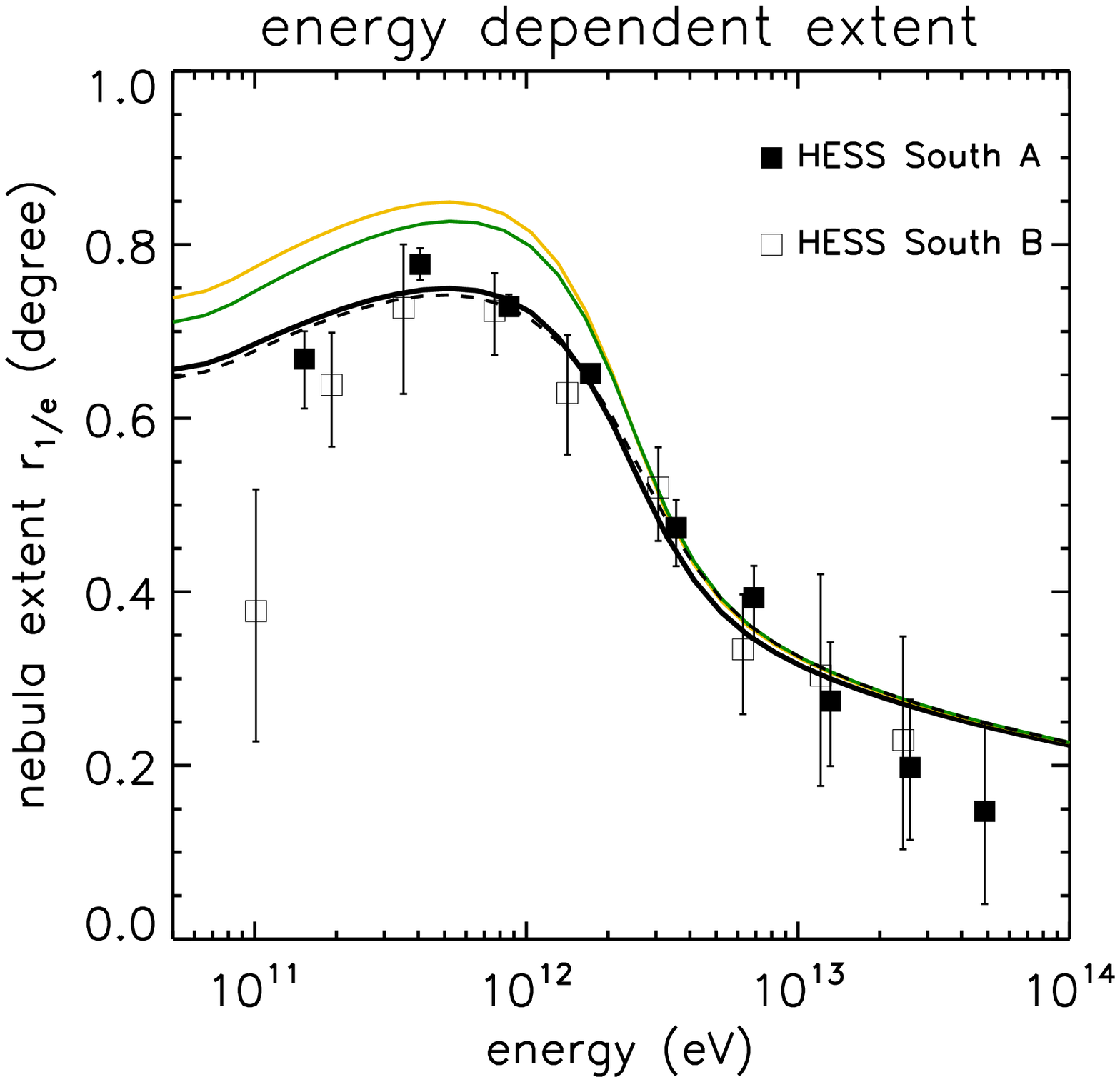}
\caption{Comparison of the predicted counts profile (left) and the energy-dependent extent (right) with different viewing angle $\phi$. Other parameters follow those in the benchmark case, except a bit larger initial rotation period of the pulsar $P_0=30\,$ms is adopted for black dashed curves.}\label{fig:extent_compare}
\end{figure*}

\bibliography{ms_v3.bib}

\end{document}